\documentstyle[aps]{revtex}

\begin{document}
\title{Reply to ``Comment on `Majoron emitting neutrinoless double beta decay in
the electroweak chiral gauge extensions' ''}
\author{S. Shelly Sharma\thanks{%
email: shelly@npd.uel.br}}
\address{Departamento de F\'{i}sica, Universidade Estadual de Londrina, Londrina,\\
Paran\'{a} 86051-970, Brazil }
\author{F. Pisano}
\address{Departamento de F\'{i}sica, Universidade Federal do Paran\'{a}, Curitiba,\\
Paran\'{a} 81531-990, Brazil }
\maketitle

\begin{abstract}
We demonstrate that in the process of deducing the constraint on the
electroweak mixing angle $\theta _{W}$ in our paper\cite{pisa98}, we have
indeed been working with three mass scales while implementing (331) model.
\end{abstract}

In their comment Montero et al. point out that there must exist at least
three different mass scales for the scalar vacuum expectation values if
majoron like scheme is implemented within the SU(3)$_{C}\otimes $SU(3)$%
_{L}\otimes $U(1)$_{X}$ model. We agree with the authors of the comment and
take this opportunity to correct a typo in our paper\cite{pisa98} as well as
demonstrate that we have in fact been working with three mass scales in the
process of deducing the constraint on the electroweak mixing angle $\theta
_{W}$.

We re-examine the generation of masses in $G_{331}\equiv \mbox{SU(3)}%
_{C}\otimes \mbox{SU(3)}_{L}\otimes \mbox{U(1)}_{X}$model. To implement the
symmetry breaking hierarchy, 
\begin{equation}
G_{331}\rightarrow G_{321}\rightarrow \mbox{SU(3)}_{C}\otimes \mbox{U(1)}_{%
{\rm em}{}},  \label{eq1}
\end{equation}
a scalar sector composed of  SU(3) symmetric sextet of scalar fields, 
\begin{equation}
S=\left( 
\begin{array}{ccc}
\sigma _{1}^{0} & h_{2}^{+} & h_{1}^{-} \\ 
h_{2}^{+} & H_{1}^{++} & \sigma _{2}^{0} \\ 
h_{1}^{-} & \sigma _{2}^{0} & H_{2}^{--}
\end{array}
\right) \sim ({\bf 1},{\bf 6}_{{\bf s}}^{*},0),  \label{eq2}
\end{equation}
and SU(3)$_{L}$ triplets, 
\begin{equation}
\eta \sim ({\bf 1},{\bf 3},0),\quad \rho \sim ({\bf 1},{\bf 3},+1),\quad
\chi \sim ({\bf 1},{\bf 3},-1),  \label{eq3}
\end{equation}
with the vacuum structure 
\begin{equation}
\langle \eta \rangle =(v_{\eta },0,0),\quad \langle \rho \rangle =(0,v_{\rho
},0),\quad \langle \chi \rangle =(0,0,v_{\chi }),  \label{eq4}
\end{equation}
and 
\begin{equation}
\langle S\rangle =\left( 
\begin{array}{ccc}
v_{\sigma _{1}} & 0 & 0 \\ 
0 & 0 & v_{\sigma _{2}} \\ 
0 & v_{\sigma _{2}} & 0
\end{array}
\right) ,  \label{eq5}
\end{equation}
is introduced. One may note that the introduction of sextet is not essential
for the symmetry breaking. In case the Eq. (\ref{eq4}) holds, the gauge
symmetry breaks to $\mbox{SU(3)}_{C}\otimes \mbox{U(1)}_{{\rm em}}$. However
it results in an antisymmetric mass matrix for charged leptons with one
eigenvalue being zero and other two equal in magnitude, for three
generations. A VEV of the sextet is needed to produce a realistic mass
matrix of the charged leptons\cite{Foot93}.

At this point we would like to correct a typo in our paper due to which it
appeared as if only two mass scales have been used. The first line after Eq.
(21) of our paper\cite{pisa98} should read

``For $v_{\sigma _{1}}=0$, notice that even if $v_{\eta }\approx v_{\rho
}\approx \sqrt{2}v_{\sigma _{2}}\equiv v_{1}$, the VEV $v_{\chi }\equiv
v_{2} $ must be large enough in order to leave the new gauge bosons
sufficiently heavy to keep consistency with low energy phenomenology. ''

instead of

``Notice that even if $v_{\eta }\approx v_{\rho }\approx v_{\sigma
_{1}}\approx v_{\sigma _{2}}\equiv v_{1}$ where $v_{1}$ denotes the usual
vacuum expectation value for the Higgs boson of the standard model, the VEV $%
v_{\chi }\equiv v_{2}$ must be large enough in order to leave the new gauge
bosons sufficiently heavy to keep consistency with low energy phenomenology.
''

We recall here that the scalar field $\sigma _{1}^{0}$ transforms as a
triplet and $\sigma _{2}^{0}$ transforms as a doublet of the subgroup SU(2).
The Masses of Charged vector bosons are

\begin{equation}
M_{W}^{2}=\frac{g^{2}}{2}\left( v_{\eta }^{2}+v_{\rho }^{2}+2v_{\sigma
_{2}}^{2}+2v_{\sigma _{1}}^{2}\right)   \label{eq6}
\end{equation}
and 
\begin{equation}
M_{Z}^{2}=\frac{g^{2}}{2c_{W}^{2}}(v_{\eta }^{2}+v_{\rho }^{2}+2v_{\sigma
_{2}}^{2}+4v_{\sigma _{1}}^{2})  \label{eq6a}
\end{equation}
An extra overall factor of $\frac{1}{2}$ and different coefficients for $%
v_{\sigma _{1}}^{2}$ in Eqs. (2) and (3) of Ref. \cite{Mont99} as compared
to our Eqs. (\ref{eq6}) and (\ref{eq6a}) above is due to a difference in
their choice of vacuum structure for SU(3)$_{L}$ triplets and symmetric
sextet of scalar fields in comparison with ours. Now with the approximation $%
v_{\eta }\approx v_{\rho }\approx \sqrt{2}v_{\sigma _{2}}\equiv $ $v_{1}$we
obtain 
\begin{equation}
M_{W}^{2}=\frac{g^{2}}{2}\left( 3v_{1}^{2}+2v_{\sigma _{1}}^{2}\right) ,
\label{eq7a}
\end{equation}
\begin{equation}
M_{Z}^{2}=\frac{g^{2}}{2c_{W}^{2}}\left( 3v_{1}^{2}+4v_{\sigma
_{1}}^{2}\right)   \label{eq7}
\end{equation}

The order of magnitude of $v_{\sigma _{1}}$ can be estimated from the
experimental constraint that is the value of $\rho $-parameter: $\rho
=0.9998\pm 0.0008$ . Using Eqs. (\ref{eq7a}) and (\ref{eq7}), we obtain 
\[
0.9998=\frac{1+\frac{2}{3}r}{1+\frac{4}{3}r}\qquad ;\text{ giving }\qquad 
\sqrt{r}=\frac{v_{\sigma _{1}}}{v_{1}}=0.0173. 
\]
For the choice $v_{1}^{2}=\left( \frac{246}{\sqrt{6}}\right) ^{2}$GeV$^{2}$ $%
\approx (100)^{2}$GeV$^{2}$,$^{\text{ }}$we get $v_{\sigma _{1}}\leq 1.73$
GeV. The following treatment leading to the constraint on the electroweak
mixing angle $\theta _{W}$ deals with a very special choice that is $%
v_{\sigma _{1}}=0$ , and $v_{2}$ $\gg v_{1}$ . For this particular case Eq. (%
\ref{eq7a}) gives $M_{W}^{2}=\frac{3}{2}g^{2}v_{1}^{2}$ and $\rho =1.$

Using the dimensionless parameters 
\begin{equation}
A\equiv \left( \frac{v_{1}}{v_{2}}\right) ^{2}  \label{eq8}
\end{equation}

and 
\begin{equation}
t\equiv \frac{g^{\prime }}{g}  \label{eq9}
\end{equation}
where $g$ and $g^{\prime }$ are the SU(3)$_{L}$ and U(1)$_{X}$ gauge
coupling constants the mass matrix for the neutral gauge bosons in the $%
\{W_{\mu }^{3},W_{\mu }^{8},B_{\mu }\}$ basis is 
\begin{equation}
\frac{1}{2}M^{2}=\frac{1}{4}g^{2}\,v_{2}^{2}\left( 
\begin{array}{ccc}
3A & \frac{1}{\sqrt{3}}A & -2tA \\ 
\frac{1}{\sqrt{3}}A & \frac{1}{3}(3A+4) & \frac{2}{\sqrt{3}}t(A+2) \\ 
-2tA & \frac{2}{\sqrt{3}}t(A+2) & 4t^{2}(A+1)
\end{array}
\right)   \label{eq10}
\end{equation}
which is a singular matrix due to the vanishing eigenvalue associated to the
photon mass. The nonvanishing eigenvalues, in the limit $A\rightarrow 0$,
are 
\begin{equation}
M_{Z}^{2}=\frac{3}{2}g^{2}\frac{1+4t^{2}}{1+3t^{2}}v_{1}^{2}  \label{eq11}
\end{equation}
for the lighter bosons and 
\begin{equation}
M_{Z^{\prime }}^{2}=\frac{2}{3}g^{2}(1+3t^{2})v_{2}^{2}  \label{eq12}
\end{equation}
for the heavier neutral Hermitian gauge boson $Z^{\prime }$. On the other
hand from Eq. (\ref{eq7a}, the counterparts of charged non-Hermitian
standard model gauge boson, have the following mass 
\begin{equation}
M_{W^{\pm }}^{2}=\frac{3}{2}g^{2}v_{1}^{2}  \label{eq13}
\end{equation}
so that in $(331)$ gauge extension 
\begin{equation}
\frac{M_{Z}^{2}}{M_{W^{\pm }}^{2}}=\frac{1+4t^{2}}{1+3t^{2}}.  \label{eq14}
\end{equation}
Comparing with the standard model result, 
\begin{equation}
\frac{M_{Z}^{2}}{M_{W^{\pm }}^{2}}=\frac{1}{1-\sin ^{2}\theta _{W}},
\label{eq15}
\end{equation}
one obtains 
\begin{equation}
t^{2}=\frac{\sin ^{2}\theta _{W}}{1-4\sin ^{2}\theta _{W}}.  \label{eq16}
\end{equation}
Therefore the theory imposes an upper bound 
\begin{equation}
\sin ^{2}\theta _{W}<\frac{1}{4}  \label{eq17}
\end{equation}
with a Landau pole in $\sin ^{2}\theta _{W}=1/4$. It is pertinent to point
out here that the limiting condition $A\rightarrow 0$ used to obtain the
constraint on the electroweak mixing angle $\theta _{W}$ in our paper
implies a $v_{\chi }\equiv v_{2}$ on TeV scale for the choice $v_{1}\sim 100$
GeV.

As such we are in fact dealing with three different mass scales represented
by scalar expectation values, $v_{1}$(chosen to be $\sim 100\,$GeV for
establishing an upper limit on $v_{\sigma _{1}}$) related with SU(3)$%
_{L}\otimes $U(1)$_{X}$ symmetry breaking, $v_{\chi }\equiv v_{2}$(on TeV
scale for the choice $v_{1}\sim 100$ GeV), large enough to leave the new
gauge bosons sufficiently heavy to keep consistency with low energy
phenomenology and $v_{\sigma _{1}}$ ( $v_{\sigma _{1}}\leq 1.73$ GeV for $%
v_{1}\sim 100$ GeV ) much smaller than $v_{1}$ consistent with the
experimental value of the $\rho $ parameter. The choice of $v_{\sigma
_{1}}=0 $ giving $\rho =1$ has been used to obtain the mass matrix for the
neutral gauge bosons in the $\{W_{\mu }^{3},W_{\mu }^{8},B_{\mu }\}$ basis
in our paper.

We may also point out that although the existence of three scales is an
important feature of the model at hand , it has no bearing on the double
beta decay related features discussed in our paper.

\acknowledgements
S. S. S. acknowledges financial support from Conselho Nacional de
Ci\^{e}ncia e Tecnologia (CNPq), Brazil.

\end{document}